\documentstyle[amssymb,preprint,aps]{revtex}

\begin{document}
\title{Anomalous doping dependence of the fluctuation-induced diamagnetism in
superconductors of YBCO family.}
\author{A.Lascialfari$^{1}$, A. Rigamonti$^{1}$ L.Romano'$^{2}$, P.Tedesco$^{1}$
,A.Varlamov$^{3}$ and D. Embriaco$^{4}$}
\address{$^{1}$ Department of Physics ''A.Volta'' and Unita' INFM, University of\\
Pavia,\\
Via Bassi n. 6, Pavia, I-27100 (Italy)\\
$^{2}$Department of Physics and Unita' INFM, University of Parma, Parco Area%
\\
delle Scienze n 7A, Parma, I-43100 (Italy)\\
$^{3}$Unita' INFM ''Tor Vergata'' , Department STFE, University of Roma\\
Tor Vergata 110, Roma, I-00133 (Italy)\\
$^{4}$Department of Physics and Unita' INFM, University of Pisa, Piazza\\
Torricelli 2, Pisa, I-56126(Italy)}


\maketitle


\begin{abstract}
SQUID magnetization measurements in oriented powders of Y$_{1-x}$Ca$_{x}$Ba$%
_{2}$Cu$_{3}$O$_{y}$, with $x$ ranging from $0$ to $0.2$, for $y\approx 6.1$
and $y\approx 6.97$, have been performed in order to study the doping
dependence of the fluctuating diamagnetism above the superconducting
transition temperature $T_{c}$. While for optimally doped compounds the
diamagnetic susceptibility and the magnetization curves $-M_{fl}(T=const$)
vs. $H$ are rather well justified on the basis of an anisotropic
Ginzburg-Landau (GL) functional, in underdoped and overdoped regimes an
anomalous diamagnetism is observed, with a large enhancement with respect to
the GL scenario. Furthermore the shape of magnetization curves differs
strongly from the one derived in that scheme. The anomalies are discussed in
terms of phase fluctuations of the order parameter in a layered system of
vortices and in the assumption of charge inhomogeneities inducing local, non
percolating, superconducting regions with $T_{c}^{(loc)}$ higher than the
resistive transition temperature $T_{c}$. The susceptibility displays
activated temperature behavior, a mark characteristic of the
vortex-antivortex description, while history dependent magnetization, with
relaxation after zero-field cooling, is consistent with the hypothesis of
superconducting droplets in the normal state. Thus the theoretical picture
consistently accounts for most experimental findings.
\end{abstract}


\vfill


\newpage


\newpage


\section{\protect\bigskip Introduction}


A variety of experiments$^{1}$ points out that the small coherence length,
reduced carrier density, high transition temperature $T_{c}$ and marked
anisotropy of cuprate superconductors cause strong enhancement of
superconducting fluctuations (SF). In contrast to conventional
superconductors, in cuprates the transition region is considerably smeared \
by SF\, which can be detected in a wide temperature range, up to 10-15 K.
The formation of the fluctuation Cooper pairs above $T_{c}$ results in the
appearance of a Langevin-type diamagnetic contribution to the magnetization $%
-M_{fl}(T,H)$, existing side by side to the paramagnetic contribution from
fermionic carriers.


Since the size of fluctuating pairs $\xi (T)$ grows when $T$ approaches the
transition temperature $T_{c}$, $M_{fl}(T,H)$ should diverge near the
transition for any small fixed magnetic field, being equal zero for $H=0.$
On the other hand it is evident that very strong magnetic fields, comparable
to $H_{c2}(0),$ must suppress SF. Therefore the isothermal magnetization
curve $M_{fl}(T=const,H)$ has to exhibit an upturn. This upturn can be
quantitatively described in the framework of the exactly solvable, for any
magnetic field, zero-dimensional model$^{2}$ (superconducting granula with
the size $\ll $ $\xi (T)$) or by means of cumbersome microscopic treatment
accounting for the short-wavelength fluctuation contribution in the $3D$
case.$^{3}$


The experiments on conventional BCS superconductors show that the
magnetization is quenched for fields as low as $\sim 10^{-2}H_{c2}(0)$ (see
Ref.$^{2})$. \ The value of the upturn field \ $H_{up}$ in the magnetization
curves can be considered inversely proportional to the coherence length.$%
^{2,4}$ This explains why in optimally doped high-temperature
superconductors, the Ginzburg-Landau (GL) picture works pretty well. Here
the coherence length is so short that the quenching of fluctuating
magnetization on increasing the magnetic field has not\ yet been observed.


\ The fluctuating magnetization of layered\ superconductors\ in the vicinity
of the transition temperature and for H$\ll H_{c2}(0),$ when the contribution
of short-wavelength fluctuations$^{3}$ is negligible, can be theoretically
described$^{5-7}$ in the framework of the GL scheme with the
Lawrence-Doniach Hamiltonian.$^{8-10}$ The fluctuating diamagnetism (FD)
turns out to be a complicated nonlinear function of temperature and magnetic
field and cannot be factorized on these variables. An important role in FD
is played by the degree of anisotropy of the electronic spectrum. All these
aspect of FD have been found to occur in optimally doped YBCO.$^{7-11}$ Also
scaling arguments$^{12}$ were found$^{13,14}$ rather well obeyed in this
compound.


In underdoped YBCO, instead, marked deviations from the behaviour expected
in the framework of GL approaches have been detected. A first qualitative
claim in this regard goes back to Kanoda et al.,$^{15}$ who noticed that in
oxygen deficient YBCO the FD in small fields was enhanced. Later on, novel
features of FD in underdoped compounds have been reported $^{8,16-20}$. In
particular, in underdoped YBCO at $T_{c}\simeq 63K$ marked enhancement of
the susceptibility for fixed $(T-T_{c})$ in a field of 0.02 Tesla was
detected above $T_{c}(0)$, with magnetization curves strongly different from
the ones in optimally doped YBCO.$^{16,18}$\ Magnetization curves have been
subsequently reported $^{19}$ in underdoped $La_{1.9}Sr_{0.1}CuO_{4}$
(LASCO). For the moment we only mention that the magnetization curves
reported by Carballeira et al.$^{19}$ in underdoped LASCO, although
indicating field-affected fluctuation-induced diamagnetism, do not exhibit
the upturn with the magnetic field as the ones in underdoped YBCO that we
will discuss later on. Finally, recent magnetization data$^{20}$ as a
function of temperature in $YBa_{2}Cu_{3}O_{6.5}$ single crystal (with
transition temperature in zero field $T_{c}(0)=45K$) indicate SF obeying to $%
2D$ scaling conditions for $H^{_>}_{^\sim} 1$ Tesla and turning to $3D$
scaling for smaller fields.


Qualitative justifications of the anomalous diamagnetism in underdoped YBCO
have been tried$^{18,21}$, essentially based on the idea of charge
inhomogeneities leading to non-percolating superconducting ''drops'' or on
the extension of the theory by Ovchinnikov et al.$^{22}$, where the
anomalous diamagnetism is related to regions having local $T_{c}$'s higher
than the resistive transition temperature. The first theoretical study
specifically aimed at the description of FD in underdoped YBCO was
undertaken by Sewer and Beck $^{23}$. In the framework of the
Lawrence-Doniach model, these authors justify the temperature and field
dependences of the magnetic susceptibility by taking into account the phase
fluctuations of the order parameter, thus arriving to a layered XY-model for
a liquid of vortices.


In this paper we address the problem of the fluctuating diamagnetism in the Y%
$_{1-x}$Ca$_{x}$Ba$_{2}$Cu$_{3}$O$_{y}$ family and of its dependence from
the number of holes, by reporting SQUID magnetization measurements in a
series of samples. Preliminary results on overdoped compounds have been
presented to a meeting and published elsewhere.$^{24}$ Since some
differences in the magnetic behaviour of chain-ordered and chain-disordered
YBCO have been noticed$^{18}$, we also attained the underdoped regime by
means of $Ca^{2+}$ for $Y^{3+}$ substitution in ideally chain-empty $%
YBa_{2}Cu_{3}O_{6}$, while for the overdoped regime the same heterovalent
substitution was performed in chain-full $YBa_{2}Cu_{3}O_{7}$.


The paper is organized as follows. In Sect.II experimental details and the
majority of the experimental results are reported. The analysis of the data
(Sect.III) is first tentatively carried out on the basis of an anisotropic
free energy GL functional, within the Gaussian approximation. The
inapplicability of such an approach for non-optimally doped compounds is
stressed. Then the theory for phase fluctuations of the order parameter in
layered liquid of vortices is revised, to properly take into account terms
neglected in the previous formulation.$^{23}$ In particular
non-reversibility and relaxation effects of the magnetization are argued to
support the picture of non-percolating, locally superconducting droplets
above the resistive transition temperature, that we interpret as phase
fluctuations of a non-zero order parameter below the local irreversibility
temperature. Thus a comprehensive description of FD in the Y$_{1-x}$Ca$_{x}$%
Ba$_{2}$Cu$_{3}$O$_{y}$ family is obtained, as it is summarized in Sect. IV.
\bigskip


\section{ Experimental details and results}


The samples of chemical composition $Y_{1-x}Ca_{x}Ba_{2}Cu_{3}O_{y}$ were
prepared by solid state reactions of oxides and carbonates in flowing oxygen
at 1000 K for about 100 hours. X-Ray diffractometry was used to check the
presence of a single phase. The oxygen stoichiometry was first estimated by
thermogravimetry and energy dispersive spectrometry. The samples were then
oxygenated close to $y=7$ by annealing in oxygen atmosphere (25 Atm) at$\ $%
450 K for about 100 hours or deoxygenated as much as possible close to $y=6$%
, for about 100 hours in vacuum. The final oxygen content turned out $%
y=6.97\pm \ 0.02$ for overdoped YBCO and $y=6.10\pm \ 0.05$ for the
underdoped samples, estimated with loss of mass measures. Before the
measurements, the samples stay at room temperature for about 1 week. The
resistive transition in samples of the same batch appeared very sharp, with
moderate evidence of paraconductivity in a temperature range of 5-10 K above
the transition. After mixing the samples with epoxy resin, they were
oriented by hardening in a strong magnetic field ( 9 Tesla). The orientation
was tested by comparing the diamagnetic susceptibility for H//c with the one
for H in the ab plane, where practically no enhancement of M$_{fl}$ was
noted to occur. For samples used in previous works $^{18,24}$, the
orientation was also tested by means of the $^{63}$Cu NMR line (see ref. 18).


Magnetization measurements have been carried out in the oriented powders by
means of a Quantum Design MPMS-XL7 SQUID magnetometer. Measurements were
performed also in optimally doped YBCO in order to prove that the results in
oriented powders do not significantly differ from the ones in single
crystals. The data already obtained by other authors$^{8,9,13,25-27}$ were
confirmed. In the following Section we will recall a few results of the
studies in optimally doped compounds, when it is required for the comparison
of our data in strongly underdoped or overdoped YBCO:Ca.


The transition temperatures $T_{c}(H=0)=T_{c}(0)$ have been estimated from
the magnetization curves vs. $T$ at small fields (20 Oersted), by
extrapolating at M=0 the linear behavior of $\chi$ occurring below T$_c$, as
shown in the insets in Fig.1. The values of $T_{c}(0)$ are collected in
Table I, where the numbers of holes $n_{h}$, as evaluated from the
expression $(\frac{T_{c}}{T_{c}^{max}})=1-82.6(n_{h}-0.16)^{2}$, giving the
parabolic behaviour$^{28}$ of the phase diagram $T-x$, are also reported. It
is noted that because of the enhanced fluctuating diamagnetism some
uncertainty in the estimate of T$_c$(0) is present, particularly in strongly
underdoped samples. This uncertainty does not affect the discussion given
later on about the anomalous diamagnetism, which is detected in a
temperature region well above T$_c$(0).


Magnetization measurements at constant field H have been performed as a
function of temperature, with ${\bf H}$ $\parallel$ ${\bf c}$. In general
two contributions to the magnetization M were observed: a Pauli-like,
positive term $M_{P}$, almost $T$-independent or only slightly increasing
with decreasing temperature in the range $\Delta T$ from 200K down to about
100 K and a negative diamagnetic contribution $-M_{fl}$ arising on
approaching $T_{c}$ . This latter contribution was extracted by subtracting
from M the value obtained by extrapolating for $T\rightarrow T_{c}^{+}$ the
curve $M_{P}$ vs. $T$ \ in $\Delta T$, where $M_{fl}$ is practically zero.
Thus the possible slight temperature dependence of $M_{P}$ around $T_{c}$
was neglected in comparison to the much stronger diamagnetic term.


Typical magnetization curves $M(H=const,T)$for overdoped and underdoped
samples are reported in Fig.2. The enhancement of FD, in both regimes, is
evidenced.


In Fig.3 some isothermal magnetization curves $-M_{fl}(T=const)$ vs H,
obtained by cooling in zero magnetic field (ZFC) down to a certain
temperature above $T_{c}(0)$, are shown. In Fig 3c also a few data for field
cooled (FC) magnetization, to be discussed later on, are reported.


\section{ Analysis of the data, further results and the theoretical picture}


\subsection{ GL anisotropic free energy functional}


The generalization of the GL functional for layered superconductors
[Lawrence-Doniach (LD) functional $^{9,11}$] in a perpendicular magnetic
field can be written 
\begin{eqnarray}
{\cal F}_{LD}\left[ \Psi \right] &=&{\sum_{l}}\int d^{2}r\left( \alpha
\left| \Psi _{l}\right| ^{2}+\frac{\beta }{2}\left| \Psi _{l}\right| ^{4}+%
\frac{h^2}{16\pi^2 m}\left| \left( {\bf \nabla }_{\shortparallel }-\frac{2ie%
}{c\hbar }{\bf A}_{\shortparallel }\right) \Psi _{l}\right| ^{2}\right. 
\nonumber \\
&&\left. +{\cal J}\left| \Psi _{l+1}-\Psi _{l}\right| ^{2}\right) ,
\label{LDF}
\end{eqnarray}
where $\Psi _{l}$\ is the order parameter of the $l-$th superconducting
layer and the phenomenological constant ${\cal J}$ is proportional to the
Josephson coupling between adjacent planes and $\alpha =\alpha _{0}(%
{\displaystyle{T-T_{c} \over T_{c}}}%
)\equiv \alpha _{0}\varepsilon $. The gauge $A_{z}=0$ is chosen in Eq.(1).
In the vicinity of $T_{c}$\ the LD functional is reduced to the GL one with
the effective mass $M=(4{\cal J}s^{2})^{-1}$ along $c$-direction,\ where $s$%
\ is the inter-layer spacing. In the GL region the fourth order term in (1)
is omitted and the standard procedure$^{2,4}$ to derive the fluctuation part
of the free energy yields


\begin{eqnarray}
&&F(\epsilon ,H)-F(\epsilon ,0)=  \eqnum{2}  \label{deltaF} \\
&=&-\frac{TVk_{B}}{2\pi s\xi _{ab}^{2}}h\int_{-\pi }^{\pi }dz{\sum_{n=0}}%
\int_{-1/2}^{1/2}dx\ln \frac{(2n+1+2x)h+r/2(1-\cos z)+\epsilon }{%
(2n+1)h+r/2(1-\cos z)+\epsilon }.  \nonumber
\end{eqnarray}
where the $c$-axis is along the z direction, $r=\frac{4\xi _{c}^{2}}{s^{2}}$
and $h=\frac{H}{H_{c2}(0)}$.


By means of numerical derivation of Eq.2 with respect to the field one
obtains the fluctuating magnetization $M_{fl}$ vs. $H$. As shown in the
inset of Fig.4, the magnetization curves in optimally doped YBCO are
satisfactorily fitted by $M_{fl}$ derived in this way and evidence how the $%
3D$ scenario of SF is obeyed on approaching $T_{c}$, with a crossover from
linear to non-linear field dependence occurring a few degrees above the
transition. Correspondingly, the scaling arguments for $3D$ anisotropic
systems hold and $M_{fl}/H^{1/2}$ vs. $T$ cross at $T_{c}(0)\simeq 92K$, as
already observed$^{13}$.


In contrast to optimally doped YBCO, the magnetization curves for underdoped
and overdoped compounds depart in a dramatic way from the ones expected on
the basis of Eq.2. In particular (see Fig.2a-3b-3c), even relatively far
from $T_{c}$, while for small fields 
(H$
(H$
linear in $H$, upon increasing the field the magnetization shows an upturn
and then $|-M_{fl}|$ decreases. Let us remind that in the GL weak
fluctuation regime the saturation of the magnetization at high field has to
be expected$^{10,11}$, the superconducting coherence being broken for fields
larger than $\sqrt{\varepsilon }H_{c2}$. An estimate of the order of
magnitude of the upturn field $H_{up}$ can be done from the analysis of the '%
$0D$' case$^{2,4}$, namely for superconducting granules of radius $d$
smaller than the coherence length $\xi (T)$. In this case the order
parameter is spatially homogeneous and the exact solution of the GL model
can be found and yields


\begin{equation}
M_{fl}=-\frac{k_{B}T\frac{2}{5}\frac{\pi ^{2}\xi ^{2}}{\Phi _{0}^{2}}d^{2}H}{%
(\varepsilon +\frac{\pi ^{2}\xi ^{2}}{5\Phi _{0}^{2}}H^{2}d^{2})}.  \eqnum{3}
\end{equation}


It can be noticed that the most sizeable contribution to the magnetization
comes from the fluctuations-induced SC droplets of radius of the order of $%
\xi (T)$ , which imply most efficient screening.$^{29}$ By assuming the
condition of zero dimension for these droplets, from Eq.3 with $d=\xi (T) $
one derives an upturn field given by $H_{up}\simeq 
{\displaystyle{\varepsilon \Phi _{0} \over \xi ^{2}}}%
$. For $\xi \simeq 10\AA $\ and $\varepsilon $ in the range $10^{-1}-10^{-2}$
, $H_{up}$ is expected to be in the range of ten Tesla.


Thus the magnetization curves in Fig.s 3 can hardly be ascribed to the
breakdown of the GL approach of the type commonly observed in BCS
superconductors$^{2}$. In other words, a description of FD based on the GL
functional in principle should be suitable in YBCO compounds for fields
smaller than several Tesla, particularly not too close to T$_c$, as in fact
it is observed in optimally doped YBCO.$^{8,9,13,26,27}$


\subsection{Phase fluctuations and superconducting droplets above T$_{c}$ :
a theoretical picture}


Then one has to look for other explanations. As already mentioned, a recent
theory has been developed by Sewer and Beck$^{23}$ at the aim to justify the
unusual magnetization curves detected in underdoped YBCO$^{18}$. The theory
assumes a frozen amplitude of the order parameter while phase fluctuations
are taken into account. As a consequence one has to deal both with thermally
activated vortex loops and field induced vortex lines. Two major conclusions
of general character can be outlined. For small field the temperature
dependence of the susceptibility is controlled by the vortex loops density $%
n_{v}$, for which


\begin{equation}
n_{v}=n_{0}exp[-E_{0}/kT]  \eqnum{4}
\end{equation}
according to the XY model. For strong fields, instead, the vortex line
elements dominate, the vortex correlations between different layers become
relevant and $M_{fl}$ only slightly increases with H and finally it
flattens. No upturn field is predicted, at least for $H<<H_{c2}$ .


In Fig.5 the data for $\chi $, defined as $(-M_{fl}/H)$, are shown to obey
rather well to Eq.4, in correspondence to $E_{0}\simeq 940K$, in agreement
with calculations yielding for the activation energy values around 10 $T_{c}$
(see Ref.$^{23}$ and references therein). $E_{0}$ turns out to depend only
little from doping, being slightly field dependent. It{\bf \ }is necessary to%
{\bf \ }mention that the temperature dependence of the susceptibility above $%
T_{c}(0)$ differs from that{\bf \ }one measured below $T_{c}$, as shown in
Fig.5.


However the magnetization curves, as the ones reported in Figs. 3 and 4,
cannot be accounted for by a theory which does not include an upturn with
the field. Furthermore similar effects are found also in overdoped samples
(see Figs.2a and 3a).


Thus we are going to consider the second aspect possibly leading to an
anomalous diamagnetism, the one related to charge inhomogeneities causing regions
where the hole density is different from the average. Evidences of
inhomogeneous structure of cuprates have been found by means of neutron and
electron diffraction$^{30-32}$, as related to stripes and lattice effects or
to local variation in the oxygen concentration, particularly near grain
boundaries. Intrinsic inhomogeneities, with spatially dependent critical
temperature have been considered as possible cause of pseudogap phenomena$%
^{33}$. A theory for high temperature superconductivity and of the pseudogap
temperature dependence based on inhomogeneous charge distribution
with site-dependent transition temperature has been recently formulated.$%
^{34}$ In particular, Ovchinnikov et al.$^{22}$ have considered the
anomalous diamagnetism above $T_{c}$ induced by non-uniform distribution of
magnetic impurities, depressing $T_{c}$ but leaving ''islands'' which become
superconductors above the resistive transition temperature. An anomalous
large diamagnetic moment results above $T_{c}$ and in this way the strong
diamagnetic susceptibility observed in overdoped Tl-based cuprates$^{35}$
could be explained. It should be stressed, however, that in this description$%
^{22}$ the magnetization is linear in the field, since the condition of
small field is implicitly assumed. Direct evidence of inhomogeneous magnetic
domains showing diamagnetic activity above T$_c$ has been obtained by Iguchi
et al.$^{36}$ by scanning SQUID microscopy in underdoped LASCO. Regions of 
few tens of $\mu m$, precursors of bulk superconductivity have been
imaged in this remarkable work.$^{36}$


In the light of the experimental findings and of the theoretical supports
outlined above we consider now as a source of diamagnetism above $T_{c}$ the
presence of locally superconducting droplets. From the volume
susceptibility, let us say at $T_{c}-5K$ (see Fig.1), one deduces that a few
percents of the total material being superconductor above the resistive
transition, could actually justify the screening effects observed as FD. A
test of this hypothesis is obtained from the search of magnetic-history
dependent effects. It is known, in fact, that in YBCO the irreversibility
temperature is not far from $T_{c}$ and therefore if the anomalous FD has to
be attributed to locally SC droplets then one should detect differences
between ZFC and field-cooled (FC) magnetization. In Fig. 6 magnetization
curves after zero field cooling and the correspondent values of $M_{fl}$
obtained at the same temperature after cooling in the presence of a given
magnetic field, are compared. Furthermore relaxation effects have been
observed. In Fig.7 it is shown how the negative magnetization depends on
time, displaying a progressive decrease from the ZFC value towards the one
measured in FC condition. The time constant for this relaxation process is
close to the one measured in the critical state$^{37}$. It can be remarked
than in underdoped chain-disordered YBCO (Fig.3c) no upturn is observed and
the ZFC and FC magnetization curves almost coincide. The explanation that
will be supported from our theoretical picture is that the magnetization
curves without hysteretic effects refer to superconducting droplets which
are above the irreversibility temperature.


One could suspect that the occurrence of superconducting droplets results
from trivial chemical inhomogeneities of the samples. As described in the
Section on experimental details many experimental checks allow us to rule
out this hypothesis. Furthermore, samples grown with different procedures
and already used by other authors, have been studied. Thus it is believed
that the inhomogeneity does not mean the presence of macroscopic parts of
the samples at different oxygen and/or calcium content, but it is rather
intrinsic, as the ones evidenced in the experiments recalled above.
Furthermore it should be remarked that the temperature dependence of the
susceptibility above the bulk transition temperature is different from the
one occurring in the superconducting state (see Fig.5).


In the following we are going to modify the theoretical description of Sewer
and Beck$^{23}$, still keeping their basic idea of phase fluctuations but
taking into account the presence of mesoscopic ''islands'' with non-zero
average order parameter amplitude that can be below or above the local
irreversibility temperature.


Let us start, as in Ref.$^{23}$, from Eq.1 by evidencing the order parameter
phase contribution


\begin{equation}
{\cal F}_{LD}\left[ \theta \right] =\frac{1}{s}{\sum_{l}}\int d^{2}r\left\{ 
{\em J}_{\shortparallel }\left( {\bf \nabla }_{\shortparallel }\theta -\frac{%
2ie}{c\hbar }{\bf A}_{\shortparallel }\right) ^{2}+{\em J}_{{\em \perp }%
}[1-\cos (\theta _{l+1}-\theta _{l})]\right\}  \eqnum{5}
\end{equation}
where ${\em J}_{\shortparallel }=$ $%
{\displaystyle{\pi \hbar ^{2}n_{h} \over 4m_{e}}}%
$ and ${\em J}_{{\em \perp }}=2\pi {\cal J}n_{h}$ are the order parameter
phase coupling constants on the plane and between planes respectively.


In this way the occurrence of superconducting droplets below the critical
temperature is assumed, where the order parameter phase can fluctuate
producing thermal excitations (vortex and antivortex pairs in 2D, vortex
loops in anisotropic model). The potential vector ${\bf A}_{\shortparallel }$
in Eq.5 describes both the magnetic field applied parallel to the $c$ - axis
and the one induced by thermal fluctuations.


By following the 2D Coulomb gas theory, at each vortex is associated an
effective charge $q_{v}=\sqrt{2\pi {\em J}_{\shortparallel }}$ \ and a
vortex-antivortex pair has an energy $E_{0}=$ $q_{v}^{2}\ln (%
{\displaystyle{r \over \xi _{ab}}}%
)$, playing the role of an activation energy and thus yielding Eq.4. In
order to refer to the anisotropic 3D model the vortex lines (or the vertical
elements of the vortex loops) are correlated along the $c$ - axis for a
length $ns$ and a correction to $q_{v}$ was found selfconsistently.


By considering, as usual, the partition function $\ \ Z=\int D\theta \exp
(-\beta {\cal F}_{LD}\left[ \theta \right] )$ with $\beta =%
{\displaystyle{1 \over k_{B}T}}%
$, the susceptibility $\chi =%
{\displaystyle{\partial M_{fl} \over \partial H}}%
$, where $M_{fl}=%
{\displaystyle{\partial F \over \partial H}}%
$, is obtained as the sum of three contributions:


\begin{equation}
\chi =\left\langle \frac{\partial ^{2}{\cal F}_{LD}}{\partial ^{2}H}%
\right\rangle ^{2}-\beta \left\langle (\frac{\partial {\cal F}_{LD}}{%
\partial H})^{2}\right\rangle +\beta \left( \left\langle \frac{\partial 
{\cal F}_{LD}}{\partial H}\right\rangle \right) ^{2}  \eqnum{6}
\end{equation}
where $\left\langle {}\right\rangle $ means the thermal average.


In the gauge $A=-yH$, $z$ being the c-axis direction, the homogeneous
susceptibility is given by 
\begin{equation}
\chi =%
\mathrel{\mathop{\lim }\limits_{q\rightarrow 0}}%
\frac{K(q)}{q^{2}},  \eqnum{7}
\end{equation}
where 
\begin{equation}
K(q)=%
{\displaystyle{J_{\shortparallel } \over d}}%
\left( \frac{2\pi }{\Phi _{0}}\right) ^{2}\left[ 
{\displaystyle{J_{\shortparallel } \over kT}}%
(P(q)-Q(q))-1\right] .  \eqnum{8}
\end{equation}


\ \ \ In Eq.(8) $P(q)$ derives from the term $\left\langle (\frac{\partial 
{\cal F}_{LD}}{\partial H})^{2}\right\rangle $ of Eq.(6) and it involves the
current-current correlation function, as in Ref.$^{23}$ :


\bigskip 
\begin{eqnarray}
P({\bf q}) &=&\frac{1}{NL^{2}}\sum_{l,l^{\prime }}\int d^{2}\rho \int
d^{2}\rho ^{\prime }\exp \left[ i{\bf q}\left( {\bf r-r}^{\prime }\right) %
\right]  \eqnum{9} \\
&&\left\langle \left( \nabla _{x}\theta _{l}(\rho )-\frac{2\pi }{\Phi _{0}}%
A_{\shortparallel ,x}\left( {\bf r}\right) \right) \left( \nabla _{x}\theta
_{l^{\prime }}(\rho ^{\prime })-\frac{2\pi }{\Phi _{0}}A_{\shortparallel
,x}\left( {\bf r}^{\prime }\right) \right) \right\rangle  \nonumber
\end{eqnarray}
with $N$ the number of layers and $L^{2}=\pi R^{2}$, $R$ being the average
radius of the superconducting islands.


The x-component of the phase gradient is


\[
\nabla _{x}\theta _{n}(\rho )=d\sum_{s_{1},l_{1}}\frac{y-R_{y}\left(
m_{1},l_{1}\right) }{|\left( {\bf \rho -R}\left( m_{1},l_{1}\right) \right)
^{2}+d^{2}\left( l-l_{1}\right) ^{2}|^{3/2}}t\left( m_{1},l_{1}\right) ,
\]
where $t\left( m_{1},l_{1}\right) =\pm 1$ and ${\bf R}(m_{1},l_{1})$ labels
the position of each ''pancake'' $m_{1}$ on the layer $l_{1}$.


Three terms are obtained by the evaluation of Eq.(9): $P_{\theta \theta }(q)$
, $P_{AA}(q)$ and $P_{\theta A}(q)$ (the two terms due to the correlation
between $\nabla _{x}\theta $ and $A_{\shortparallel ,x}$ give the same
contribution being $P_{\theta A}(q)=P_{A\theta }(q)=P_{\theta A}(-q)$). The
first one involves the positional correlation function of the vortex line
elements. In order to calculate it, Sewer and Beck$^{23}$ introduced the\
static structure factor of a disordered vortex liquid. Because of the weak
interlayer coupling harmonic deviations of \ the vortex lines (or loops)
along the $z$ direction are taken into account. This model can be used to
describe also the vortex system in the glassy phase, below the
irreversibility line \ temperature and therefore the same expression for $%
P_{\theta \theta }(q)$ is used here.


The evaluation of the term $P_{AA}(q)$ is straightforward and one has


\begin{equation}
P_{AA}(q)=\frac{\pi ^{2}}{36}(\frac{HL^{2}}{\Phi _{0}})L^{2}q^{2}  \eqnum{10}
\end{equation}


The further contribution $P_{\theta A}(q)$ , appearing due to the cross
correlation between $\nabla _{x}\theta ${\bf \ }and{\bf \ }$%
A_{\shortparallel ,x}${\bf \ }and disregarded in Ref{\bf .}$^{23}${\bf ,}
cannot be neglected below the vortex lattice melting temperature, where
irreversibility effects occur. In this case one obtains 
\begin{eqnarray*}
P_{\theta A}(q)+P_{\theta A}(-q) &=&2\frac{Hd}{L}\frac{\left( 2\pi \right)
^{2}}{\Phi _{0}}\frac{Lq\cos \frac{Lq}{2}-2\sin \frac{Lq}{2}}{q^{2}}%
\sum_{l,l^{\prime }}\exp \left( -dq\left| l-l_{1}\right| \right) \\
&&\left\langle \sum_{m_{1},l_{1}}t\left( m_{1},l_{1}\right) \cos \left[
iqlR_{y}\left( m_{1},l_{1}\right) \right] \right\rangle .
\end{eqnarray*}


The thermal average is performed in the assumption that the vortices are
uniformly distributed in the planes and the calculations are reported in
Appendix. The expansion of $P_{\theta A}(q)$ in powers of $qL$ gives


\begin{equation}
P_{\theta A }=-\frac{2\pi ^{2}}{3}\left( \frac{HL^{2}}{\Phi _{0}}\right)
^{2}+\frac{2L^{2}}{45}\pi ^{2}\left( \frac{HL^{2}}{\Phi _{0}}\right) 
\eqnum{11}
\end{equation}


The function $Q(q)$ in Eq.8, related to the third term of Eq.6, has been
neglected in Ref.$^{23}$. It can be calculated as described in Appendix,
yielding


\begin{equation}
Q=(2\pi )^{2}\left( \frac{HL^{2}}{\Phi _{0}}\right) ^{2}\left[ \frac{1}{%
q^{2}L^{2}}+\frac{1}{144\times 4}q^{2}L^{2}+\frac{1}{12}\right] .  \eqnum{12}
\end{equation}
It should be noted that the first term in $Q$, diverging for $q\rightarrow 0$%
, exactly cancels out the $q^{-2}$ term in the expansion of $P(q)$ which
appears from the structure factor.


By using Eqs. (10), (11), (12) and Eq.(6) of Ref.$^{23}$, from Eq.(8) one
finally obtains


\begin{eqnarray}
K(q) &=&%
{\displaystyle{J_{\shortparallel } \over s}}%
\left( \frac{2\pi }{\Phi _{0}}\right) ^{2}\left[ \frac{2\pi
J_{\shortparallel }}{q_{v}^{2}}(1+2n)-\delta \left( \frac{H}{H^{\ast }}%
\right) ^{2}-1\right] +  \eqnum{13} \\
&&\left[ -\frac{kT}{s\Phi _{0}^{2}}\frac{1}{1+2n}\frac{(1+\delta \left( 
\frac{H}{H^{\ast }}\right) ^{2})^{2}}{n_{v}}-\frac{s^{2}\gamma ^{2}(1+n)}{%
1+2n}(1+\delta \left( \frac{H}{H^{\ast }}\right) ^{2})+\right.  \nonumber \\
&&\left. \frac{47\pi R^{2}}{540}\frac{J_{\shortparallel }}{s}\left( \frac{%
2\pi }{\Phi _{0}}\right) ^{2}\delta \left( \frac{H}{H^{\ast }}\right) ^{2}%
\right] q^{2}  \nonumber
\end{eqnarray}
with $\delta =%
{\displaystyle{3\pi ^{2} \over 4}}%
{\displaystyle{J_{\shortparallel } \over kT}}%
$ and $H^{\ast }=%
{\displaystyle{\Phi _{0} \over \pi R^{2}}}%
$ is an effective ''critical'' field depending on the island size.


To avoid unphysical divergences in the calculation of the susceptibility
from Eq.(7), the first term in square brackets of Eq.(13) has to be zero,
giving a renormalization of $q_{v}$ due to both the anisotropy of the system
and the presence of applied magnetic field: 
\begin{equation}
q_{v}^{2}(H)=\frac{q_{v}^{2}(1+2n)}{(1+\delta \left( \frac{H}{H^{\ast }}%
\right) ^{2})}.  \eqnum{14}
\end{equation}
\ \ \ \ \ \ In view of the field-dependent vortex charge, the pair energy
(in the limit $H<H^{\ast} $ ) becomes $E=%
{\displaystyle{E_{0} \over (1+\delta \left( \frac{H}{H^{\ast }}\right) ^{2})}}%
$. According to Eq.(4) the thermally-excited vortex pair density turns out
field dependent. This field dependence, formally derived in our description,
is significantly different from the one assumed in Ref.$^{23}$.


Finally the diamagnetic susceptibility is obtained in the form 
\begin{equation}
\chi =-\frac{kT}{s\Phi _{0}^{2}}\frac{1}{1+2n}\frac{(1+\delta \left( \frac{H%
}{H_{\ast }}\right) ^{2})^{2}}{n_{v}}-\frac{s^{2}\gamma ^{2}(1+n)}{1+2n}%
(1+\delta \left( \frac{H}{H^{\ast }}\right) ^{2})+\frac{47\pi R^{2}}{540}%
\frac{J_{\shortparallel }}{s}\left( \frac{2\pi }{\Phi _{0}}\right)
^{2}\delta \left( \frac{H}{H^{\ast }}\right) ^{2}  \eqnum{15}
\end{equation}
In the limit $H\rightarrow 0$ a good agreement of the susceptibility and its
temperature dependence with the experimental findings is again achieved. The
main differences between our susceptibility in Eq.(15) and the one given in
Ref.$^{23}$ consists in the presence of the factor $(%
{\displaystyle{H \over H^{\ast }}}%
)^{2}$ and of the third, positive term. This term can give an inversion in
the sign of the susceptibility corresponding to an upturn in the
magnetization curves. This phenomenon depends on the dimension of the
islands and $\chi =0$ (i.e. the occurrence of the upturn) requires $R>R_{0}$
where $R_{0}$ depends on some characteristics of the material. By choosing $%
\gamma =6$, the interlayer distance $s=12$\AA , $n=2$, $%
{\displaystyle{J_{\shortparallel } \over kT}}%
=2.5$, which are typical values for YBCO, for $T=75.5K$ one estimates $%
R_{0}\simeq 50$\AA . In this case the solutions of the Equation $\chi =0$ is 
$%
{\displaystyle{H_{up} \over H^{\ast }}}%
\simeq 0.06$ and by considering the experimental value $H_{up}\simeq 250$ G,
the effective critical field turns out $H^{\ast }\simeq 0.4T$.


The isothermal curves can be obtained from Eq.(15) by means of numerical
integration. The shape of the magnetization curve depends on the parameters
in the susceptibility and by using the values quoted before, with $R=370$%
\AA\, one derives the behaviour sketched in Figs.3a for an island below the
irreversibility line. The same parameters with $T=66K$, $%
{\displaystyle{J_{\shortparallel } \over kT}}%
=1.8$ and $R=10$\AA\ lead to the curve shown in Fig.3c for the magnetization
of the island above irreversibility.


Finally we discuss the differences observed in the magnetization curves
between chain-ordered and chain-disordered YBCO compounds and the relevant
observation by Carballeira et al.$^{19}$ of magnetization curves -M$_{fl}$
vs H in underdoped LASCO (Sr content $x$=0.1, T$_c$(0)=27.1 K) with no upturn
field. A role of the chains on favouring the nucleation of local
superconducting droplets above T$_c$ is conceivable. In fact, in
chain-ordered compounds the droplets appear to have the irreversibility
temperature higher than the ones in chain-disordered compounds, as it is
evidenced by the difference in the magnetization curves (see Figs. 3a-b-c).
We remind that the inversion in the sign of the susceptibility is related to
the third term in Eq.15 and thus to the term $P_{\theta A}(q)$. The amount
of impurities and/or imperfections acting as nucleation centers might also
play a role. Furthermore the degree of under or over-doping is also involved
since a marked variation of T$_c$ with n$_h$ is evidently crucial. It is
noted that in the measurements by Carballeira et al.$^{19}$ in LASCO at T$_c$%
=27.1 K the magnetization curves show only a weak tendency to saturation
while in LASCO at T$_c \simeq$18 K (therefore more underdoped) scanning
SQUID microscopy by Iguchi et al$^{36}$ does evidence diamagnetic effects to
associate to locally superconducting droplets. These droplets should imply a
contribution to the magnetization curves similar to the one detected by us
in YBCO compounds. Future research work will have to explore these
interesting aspects and the differences until now present between LASCO and
YBCO. \newpage


\section{Conclusions}


By means of SQUID measurements in $Y_{1-x}Ca_{x}Ba_{2}Cu_{3}O_{y}$ family a
non conventional fluctuating diamagnetism has been observed in overdoped and
in underdoped compounds. Compared to optimally doped YBCO, a large
enhancement of the diamagnetic susceptibility occurs and no anisotropic GL
functional or scaling arguments can justify the isothermal magnetization
curves. The recent theory$^{23}$ for phase fluctuations of the order
parameter in a layered liquid of vortices has been revised and it appears to
justify some aspects of the anomalous FD in non-optimally doped YBCO,
particularly the ''precritical'' temperature activated behaviour of the
susceptibility in the limit of zero field. Other experimental observations,
noticeably the upturn in the field dependence of the isothermal fluctuating
magnetization and history-dependent effects, indicate the role of mesoscopic
charge inhomogeneities in inducing local, non-percolating, superconducting
''droplets''. On the basis of both types of experimental findings we have
extended the theory of phase fluctuations in the presence of non-zero order
parameter. The terms leading to a novel and relevant dependence of the
fluctuating magnetization from the magnetic field were included in the
scheme. The field-related corrections are different when the superconducting
droplets are below or above the local irreversibility temperature. In this
way most of the experimental findings have been justified.


\bigskip


\section{ACKNOWLEDGMENTS}


Thanks are due to F.Borsa, F. Licci and S. Kramer for having provided some
of the samples used in the present work and for useful discussions. F.
Cordero is thanked for the deoxygenation of the underdoped samples and
S.Sanna for the resistivity measurements as a function of temperature. One
of the authors (A.V.) thanks H.Beck and A.Sewer for interesting discussions
during his visit to Neuchatel University. D.E. has carried out the work in
the framework of the stage program of the INFM National School ( 2000).


\section{REFERENCES}


$^{1}$ See ''Fluctuation Phenomena in High Temperature Superconductors'',
Edited by M.Ausloos and A.A.Varlamov, Kluwer Academic Publishers,
Netherlands, 1997; see also A.A.Varlamov, G.Balestrino, E.Milani and
D.V.Livanov, Adv. Phys. 48, 655 (1999).


$^{2\text{ }}$M.Tinkham ''Introduction to Superconductivity'' Mc Graw-Hill,
New York 1996, Chapter 8.


$^{3}$ J.Kurkijarvi,V.Ambegaokar and G.Eilenberger, Phys.Rev B 5, 868 (1972).


$^{4}$ See A.I. Larkin and A.A.Varlamov, ''Fluctuation Phenomena in
Superconductors'' in ''Physics of conventional and non-conventional
superconductors'' Eds. K.-H.Bennemann and J.B. Ketterson , Springer Verlag
(2001).


$^{5}$ L.N. Bulaevskii, Int. J. of Modern Phys. B 4, 1849 (1990).


$^{6}$ L.N. Bulaevskii, M.Ledvig and V.G. Kogan, Phys. Rev. Lett. 68, 3773
(1993).


$^{7}$ V.G. Kogan, M. Ledvij, A. Yu. Simonov, J. H. Cho, and D. C. Johnston,
Phys. Rev. Lett. 70, 1870 (1993).


$^{8}$ M.A. Hubbard, M.B. Salamon and B.W. Veal, Physica C 259, 309 (1995).


$^{9}$ C.Baraduc, A. Buzdin, J.Y. Henry, J.P. Brison and L. Puech, Physica C
248, 138 (1995).


$^{10}$ A.E.Koshelev, Phys. Rev B 50, 506 (1994).


$^{11}$ A.Budzin and V.Dorin, in ''Fluctuation Phenomena in High Temperature
Superconductors'', Ref.1, p.335. A detailed theoretical analisys of the
field dependence of the Gaussian fluctuations in layered superconductors has
been given by T.Mishonov and E. Penev, Intern. J.of Modern Phys. B 14, 3831
(2000).


$^{12}$ T. Schneider and U. Keller, Inter. J. of Modern Phys. B 8, 487
(1993) and references therein; T. Schneider and U. Keller,Physica C 207, 336
(1993); see also T.Schneider and J.M. Singer, in ''Phase Transitions
Approach to High Temperature Superconductivity'', Imperial College Press
2000, Chapter 6.


$^{13}$ A.Junod, J-Y Genoud, G.Triscone and T.Schneider, Physica C 294, 115
(1998).


$^{14}$ S.Salem-Sugui and E.Z. Dasilva, Physica C 235, 1919 (1994).


$^{15}$ K.Kanoda, T. Kawagoe, M. Hasumi, T. Takahashi, S. Kagoshima and T.
Mizoguchi, J.Phys. Soc. Japan 57, 1554 (1988).


$^{16}$ P.Carretta, A.Lascialfari, A.Rigamonti, A.Rosso and A.A.Varlamov,
Inter. J. of Modern Phys. B 13, 1123 (1999).


$^{17}$ D.Babic,J.R.Cooper, J.W. Hodby and Chen ChangKang, Phys. Rev. B 60,
698 (1999).


$^{18}$ P.Carretta, A.Lascialfari, A.Rigamonti, A.Rosso and A.A.Varlamov,
Phys. Rev.B 61, 12420 (2000).


$^{19}$ C.Carballeira, J. Mosqueria, A.Revcolevschi and F.Vidal, Phys. Rev.
Lett. 84, 3157 (2000).


$^{20}$ B.Rosenstein, B.Y. Shapiro, R. Prozorov, A. Shaulov, Y. Yeshurun,
Phys.Rev. B 63, 134501 (2001).


$^{21}$ A.Rigamonti and P. Tedesco, Lecture Notes at the INFM National
School in Structure of Matter (Villa Gualino, Torino, September 2000) and
''Scientifica Acta'', University of Pavia, XV, 49 (2000).


$^{22}$ Yu.N. Ovchinnikov, S.A.Wolf and V.Z. Kresin, Phys.Rev. B 60, 4329
(1999).


$^{23}$ A.Sewer and H.Beck, Phys. Rev.B 64, 014510 ( 2001).


$^{24}$ P.Carretta, A Lascialfari, A.Rigamonti, F.Teodoli, F.Bolzoni e
F.Licci, Inter. J.of Modern Phys. B 14, 2815 (2000).


$^{25}$ U.Welp, S. Fleshler, W. K. Kwok, R. A. Klemm, V. M. Vinokur, J.
Downey, B. Veal, and G.  W. Crabtree, Phys.Rev. Lett. 67, 3180 (1991).


$^{26}$ C. Torron, A. Diaz, A. Pomar, J.A. Veira and F. Vidal, Phys. Rev.
B49, 13143 (1994)


$^{27}$ J. Mosqueira, C. Carballeira, M.V. Ramallo, C. Torron, J. Veira and
F. Vidal, Europhys. Lett. 53, 632 (2001)


$^{28}$ J.L. Tallon, C. Bernhard H. Shaked R. L. Hitterman and J. D.
Jorgensen, Phys. Rev. B 51 , 12911 ( 1995).


$^{29}$ J.P.Gollub, M. R. Beasley, R. Callarotti and M. Tinkham, Phys. Rev.
B 7, 3039 (1973).


$^{30}$ M.Gutmann, S.J.L. Billinge, E.L. Brosha and G.H. Kwei , Phys. Rev. B
61, 11762 (2000).


$^{31}$ E.Bozin. G.Kwei, H.Takagi and S.Billinge, Phys.Rev. Lett. 84, 5856
(2000).


$^{32}$ Z.Akase, Y.Tomokiyo, Y.Tanaka and M.Watanabe, Physica C 339, 1
(2000).


$^{33}$ Yu.N. Ovchinnikov, S.A.Wolf and V.Z. Kresin, Phys.Rev. B 63, 064524
(2001).


$^{34}$ E. V. L. De Mello, E.S. Caixeiro and J.L. Gonzales, cond-mat 0110519
(24 Oct 2001); ibidem, cond-mat 0110479 (22 Oct 2001)


$^{35}$ C.Bergemann, A. W. Tyler, A. P. Mackenzie, J. R. Cooper, S. R.
Julian and D. E. Farrell, Phys. Rev. B 57, 14387 (1998).


$^{36}$ I. Iguchi, T. Yamaguchi and A. Sugimoto, Nature 412, 420 (2001)


$^{37}$Y.Yeshurun, A.P. Malozemoff, A. Shaulov, Rev. of Modern Phys. 68, 911
(1996).


\newpage


\section{\protect\bigskip Appendix}


\bigskip To derive Eqs.(11)\ and\ (12) the following thermal average must be
calculated:


\begin{equation}
\left\langle \sum_{m,l}t\left( m,l\right) \cos iqlR_{y}\left( m,l\right)
\right\rangle =\sum_{m}\left\langle t(m)\right\rangle -\frac{1}{2}%
q^{2}\sum_{m}<R^{2}(m)>+o(q^{3})  \eqnum{A1}
\end{equation}
Indicating with $N_{+}$ ($N_{-}$) the number of vortex line elements
parallel (antiparallel) to the field the first term gives


\[
N_{+}-N_{-}=\frac{HL^{2}}{\phi _{0}}. 
\]


The sum in the second term can be split in two parts which separately count
the vortex line elements parallel and antiparallel to the field: 
\[
\sum_{m}t(m)<R_{y}^{2}(m)>=\sum_{m+}<R_{y}^{2}(m)>-\sum_{m-}<R_{y}^{2}(m)>,
\]
One can assume that the vortices are uniformely distributed in the planes
and that the $y$ components of their positions are distributed on a line,
separated each other by a distance $\Delta L=\frac{L}{N_{+}}$.Then, \ the $i-
$th vortex is in the mean position $<R_{i}>=\Delta Li=\frac{\Delta L}{N_{+}}i
$, with $i=-\frac{N_{+}}{2}\cdots \frac{N_{+}}{2}$. 
\[
\sum_{m_{\pm }}<R(m_{\pm })^{2}>=\sum_{i=-\frac{N_{+}}{2}}^{\frac{N_{+}}{2}}%
\frac{L^{2}}{N_{\pm }^{2}}i^{2}=\frac{L^{2}}{12N_{\pm }}\left( N_{\pm
}+2\right) \left( N_{\pm }+1\right) .
\]
By considering $N_{\pm }>>1${\bf \ }one finally finds 
\[
\sum_{m}t(m)<R_{y}^{2}(m)>\approx \frac{L^{2}}{12}(N_{+}-N_{-}).
\]
Then (A1) can be written 
\[
\sum_{m}t(m)<\cos {qR(m)}>=\frac{HL^{2}}{\Phi _{0}}-q^{2}\frac{L^{4}H}{%
24\Phi _{0}}+o(q^{3})
\]


\bigskip


\bigskip


\newpage


\section{FIGURE CAPTIONS}


Fig.1 Some magnetization data in low field, parallel to the c-axis, as a
function of temperature in oriented powders of $%
Y_{1-x}Ca_{x}Ba_{2}Cu_{3}O_{y}$. The values of the magnetization measured
from 250 K down to 90 K, with a positive Pauli-like temperature independent
term, are not reported in the figure. In the insets the blow-up for the
estimate of $T_{c}(0)$ is shown.


Fig.2 Constant field (${\bf {H} \parallel c}$) magnetization vs. temperature
in overdoped ($x=0.1$ and $y=6.96$) (a) and in underdoped ($x=0$ and $y=6.65$%
) (b) YBCO compounds. For comparison in the part a) of the figure the
behaviour of $M_{fl}$ in optimally doped YBCO is shown.


Fig.3 Isothermal diamagnetic magnetization vs. $H$, after zero-field cooling
(ZFC) to a certain temperature above $T_{c}(0).$ a) sample at $x=0.1$, $%
y=6.96$ ( $T_{c}(0)=73K$); b) sample at $x=0.2$, $y=6.98$ ( $T_{c}(0)=49.5K$%
); c) chain disordered underdoped YBCO at $y=6.65$ and $x=0$ ( $%
T_{c}(0)=62.5K$) for ZFC (circle) and field-cooled (FC) (up-triangle)
conditions. The solid lines in part a) of the Figure correspond to the
diamagnetic susceptibility estimated in the limit of zero field. The solid
lines in Figs.3a and 3c are the theoretical behaviours according the
mechanisms described in the text for droplets below and above the local
irreversibility temperature.


Fig.4 Comparison of the magnetization curves $M_{fl}$ vs. $H$ (after ZFC) in
overdoped YBCO:Ca ($x=0.1$, $y=6.96$) with the ones in optimally doped YBCO
(inset), for similar reduced temperature $\varepsilon $. The solid lines
fitting the data in optimally doped YBCO are derived from Eq.2 in the text
by means of numerical derivation, correspond to anisotropic parameter r=0.1
and evidence the 3D linear and 3D non-linear regimes.


Fig.5 Susceptibility as a function of the inverse temperature showing the
activated temperature behavior in the sample at $y=6.96$ and $x=0.1$.
Analogous temperature dependence has been observed in underdoped YBCO
compounds. The dashed lines are obtained by transferring above the
superconducting temperature $T_{c}(0)$ the temperature behavior of the bulk
susceptibility measured below $T_{c}$ and by normalizing the data at $%
T\simeq 88K$.


Fig.6 Magnetization curves in YBCO:Ca at $x=0.1$ and $y=6.96$ by cooling in
zero field to a given temperature [a) $T=75.5K$; b) $T=79.5K$] above the
resistive transition temperature and then applying the field (ZFC) and after
the application of the field at room temperature, cooling at the same
temperature and measuring the correspondent magnetization (FC). The volume
susceptibility in the limit H$\rightarrow $0 is reported.


Fig. 7 Relaxation of the raw magnetization after ZFC and then sudden
application of a magnetic field of $260$ G, in YBCO:Ca at $x=0.1$ and $y=6.96
$, at $T=75.5K$ a) short-term relaxation; b) long-term relaxation. From the
comparison of the ZFC and FC magnetization in H=20G (see inset) an
irreversibility temperature of the locally superconducting droplets at
highest $T_{c}$ can be estimated around 90 K. In part a) of the Figure the
solid line is the sketchy behavior of the relaxation of the magnetization
detected by Yeshurun et al. (Ref.$^{33}$) in the critical state, in
optimally doped YBCO.


\bigskip


\bigskip


\bigskip


\bigskip


\bigskip


\bigskip


\bigskip


\bigskip


\bigskip


\bigskip \newpage


\section{\protect\bigskip TABLE CAPTION}


Table I: Superconducting transition temperature in overdoped and underdoped Y%
$_{1-x}$Ca$_{x}$Ba$_{2}$Cu$_{3}$O$_{y}$ and estimated number of holes per CuO%
$_{2}$ unit


\bigskip


\bigskip {\bf TABLE I }


\bigskip $
\begin{array}{cccc}
x & y & T_{c}(K) & n_{h} \\ 
0 & 6.65 & 62.5 & 0.12 \\ 
0.05 & 6.97 & 82.0 & 0.18 \\ 
0.1 & 6.96 & 73.0 & 0.20 \\ 
0.1 & 6.96 & 70.0 & 0.21 \\ 
0.2 & 6.98 & 49.5 & 0.23 \\ 
0.15 & 6.10 & 34.00\pm 1 & 0.07 \\ 
\approx 0.15 & 6.05 & 20.00\pm 2 & 0.06 \\ 
0.1 & 6.10 & 14.00\pm 2 & 0.06
\end{array}
$


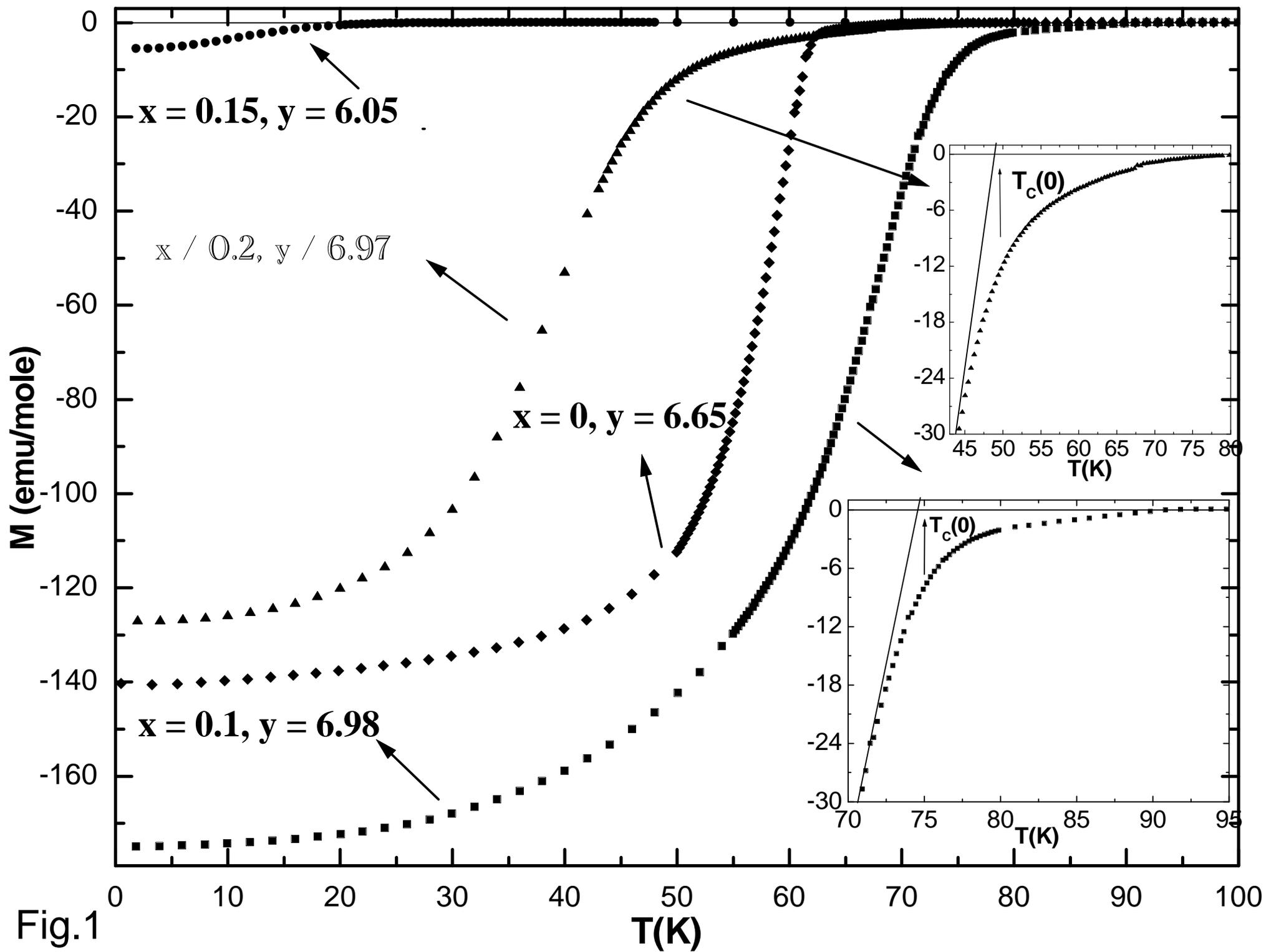
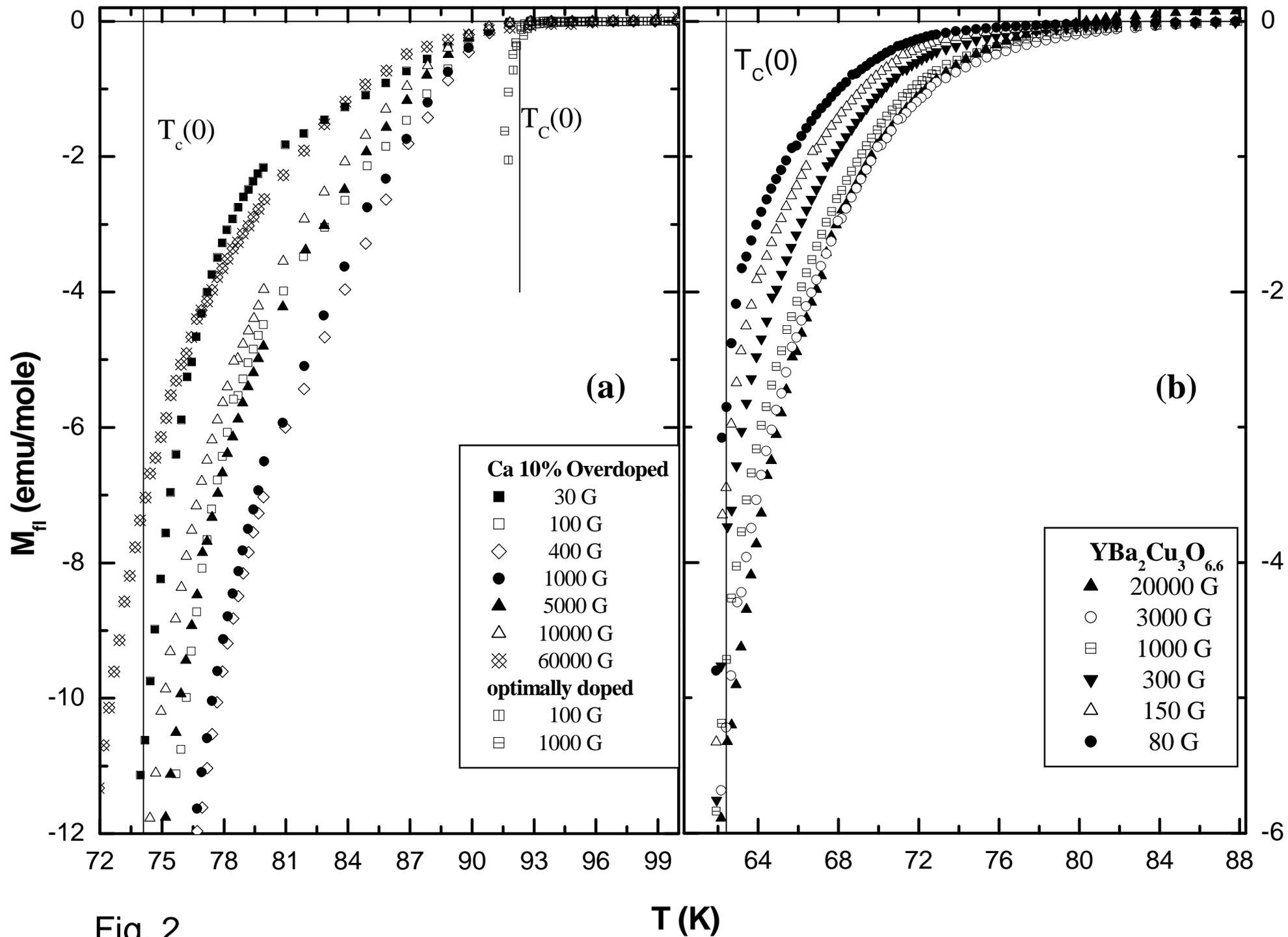
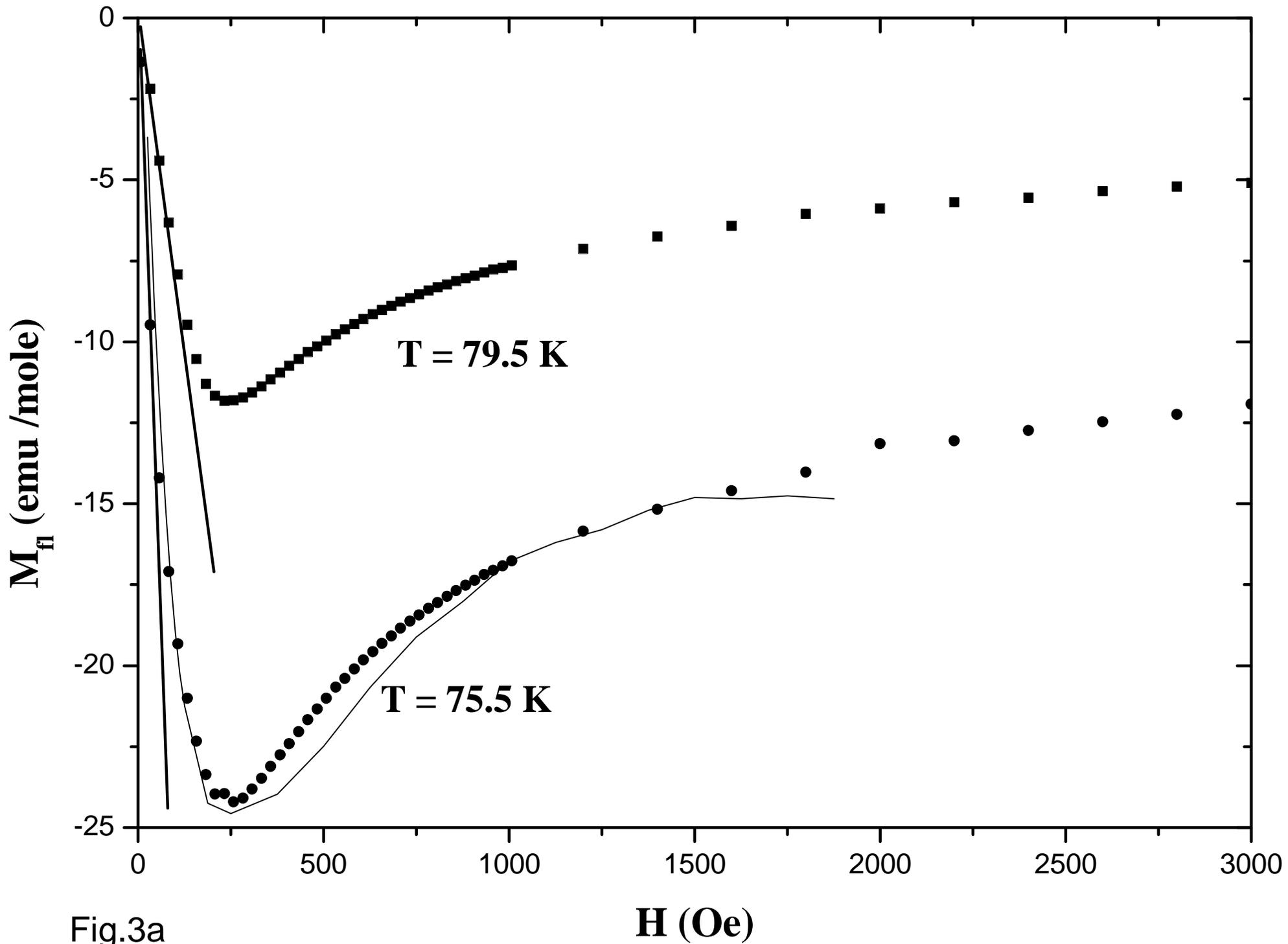
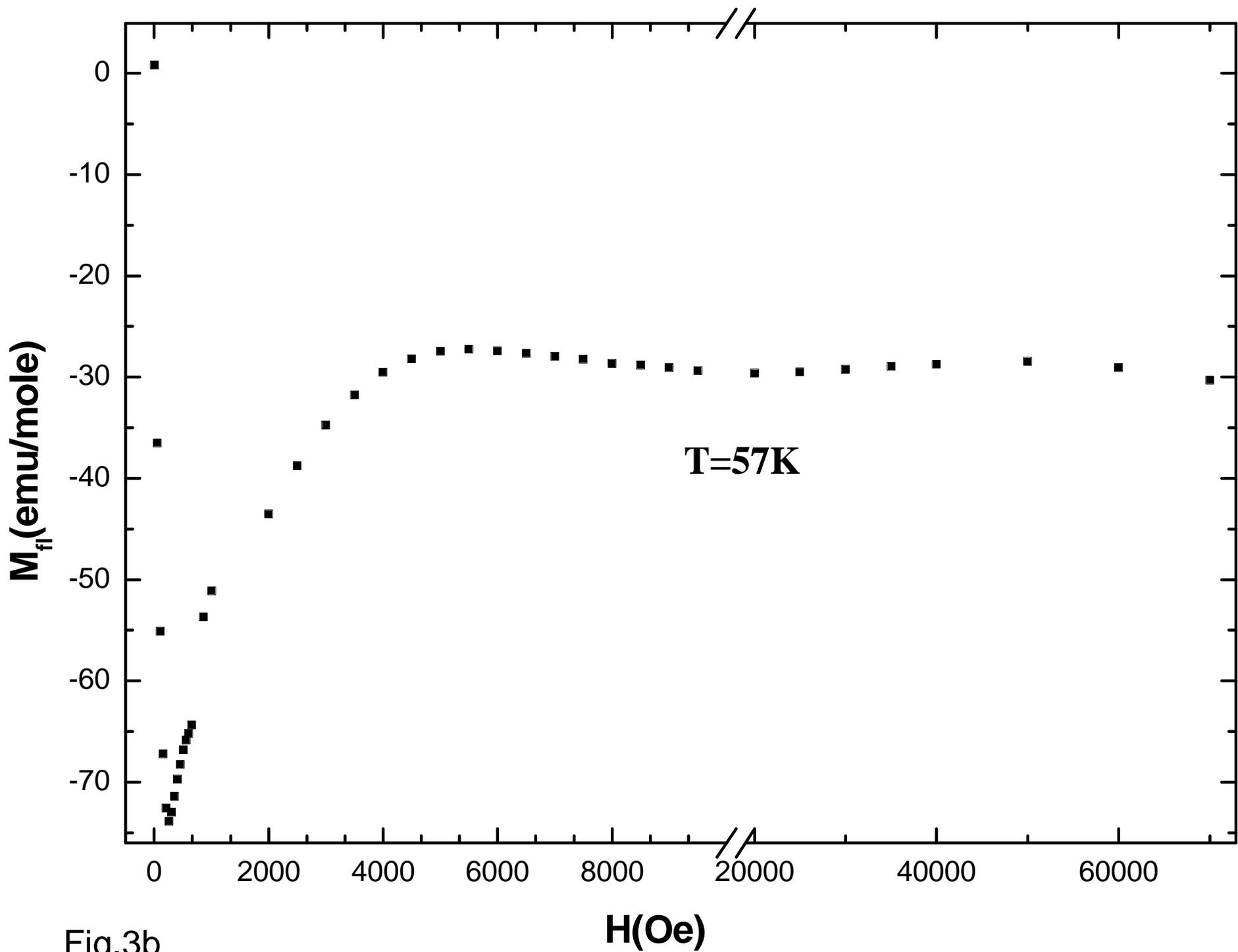
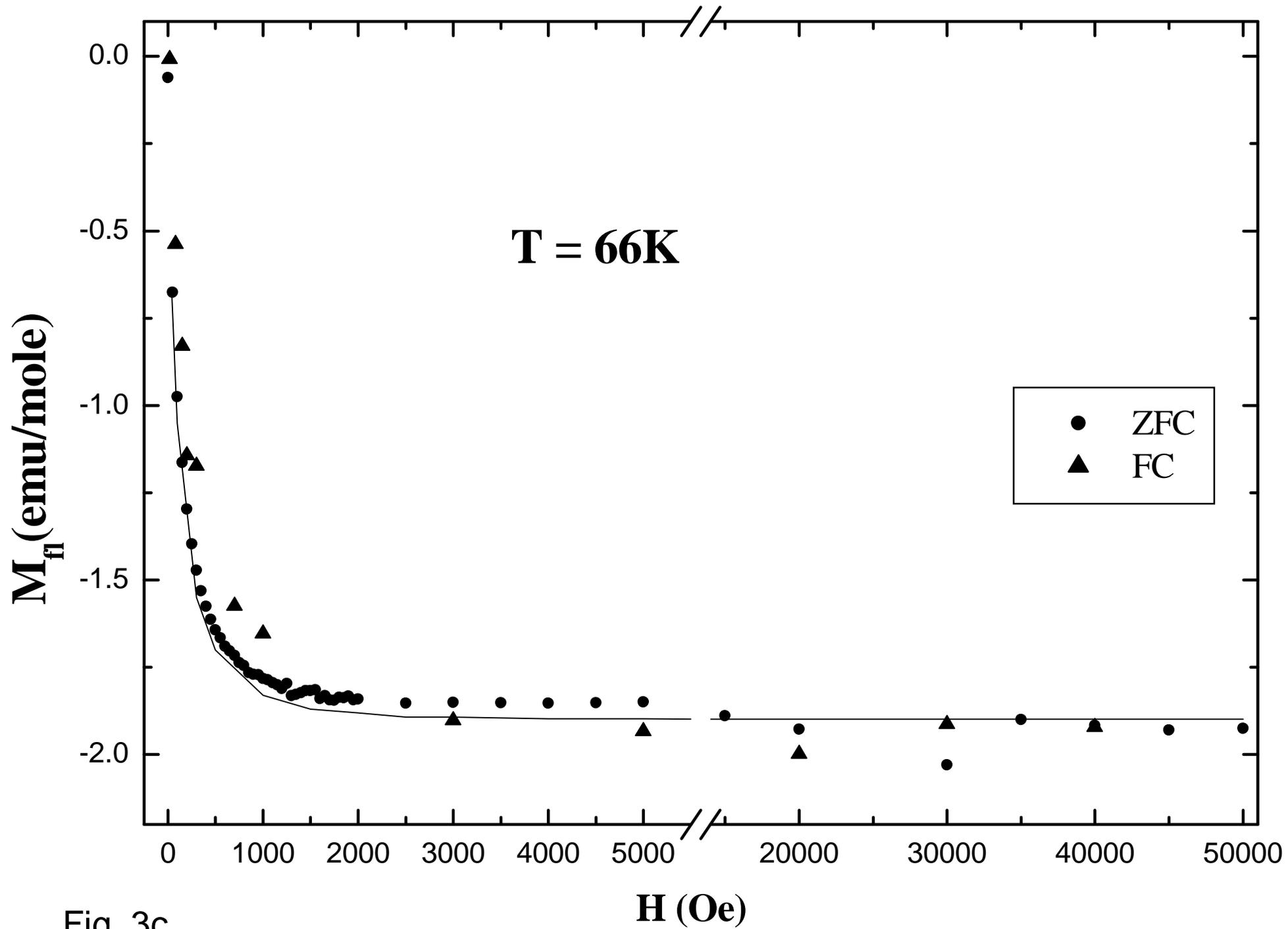
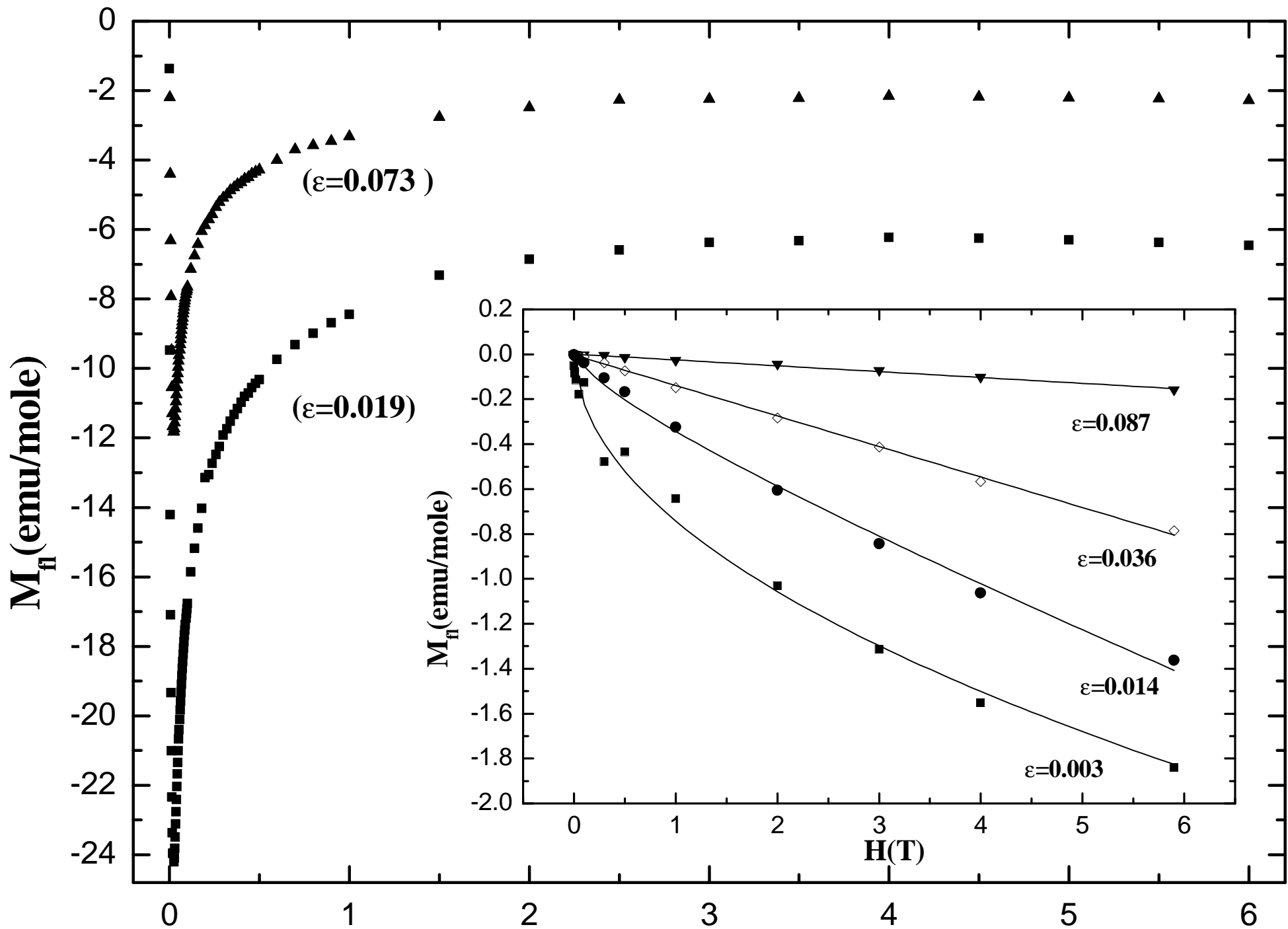
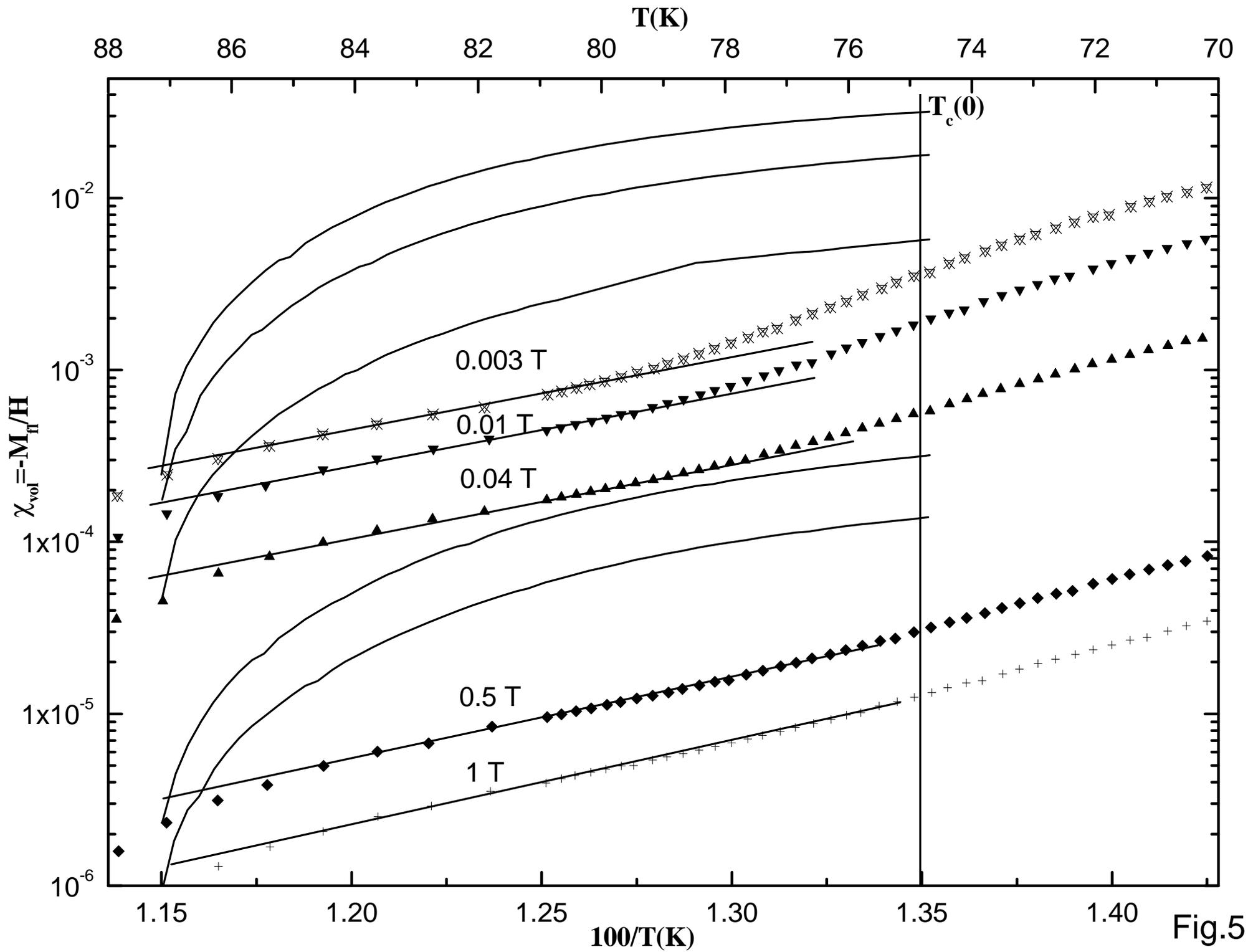
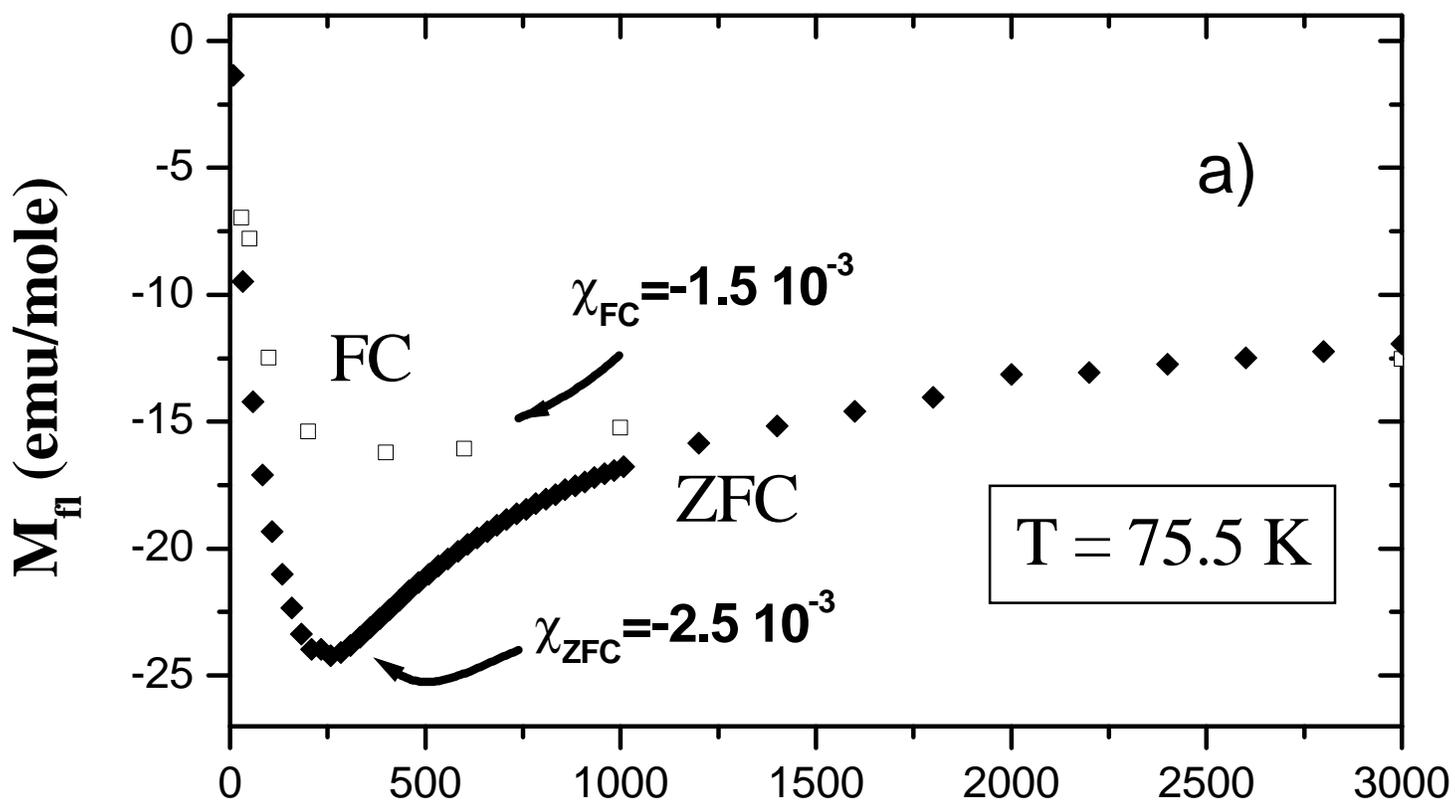
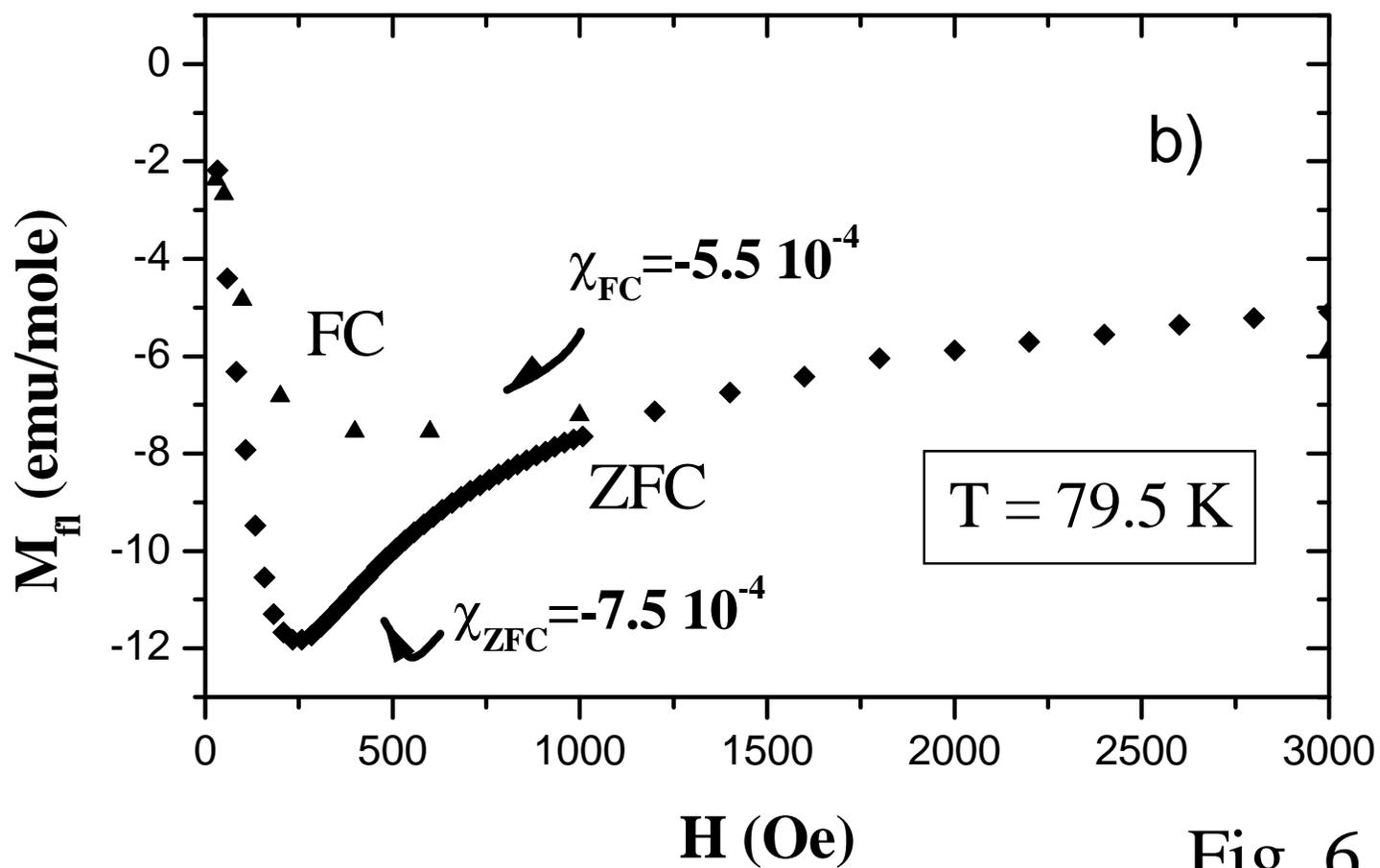
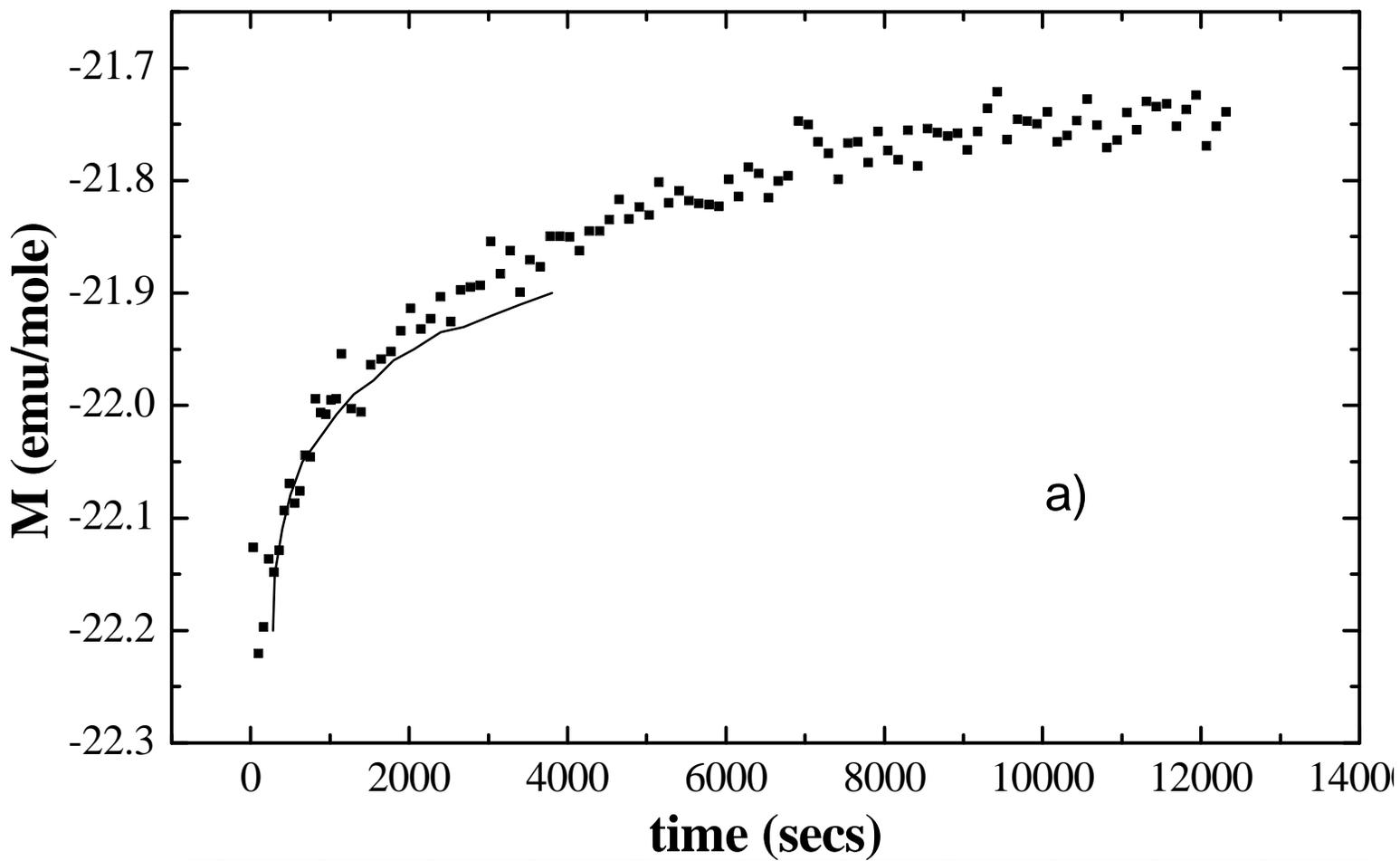
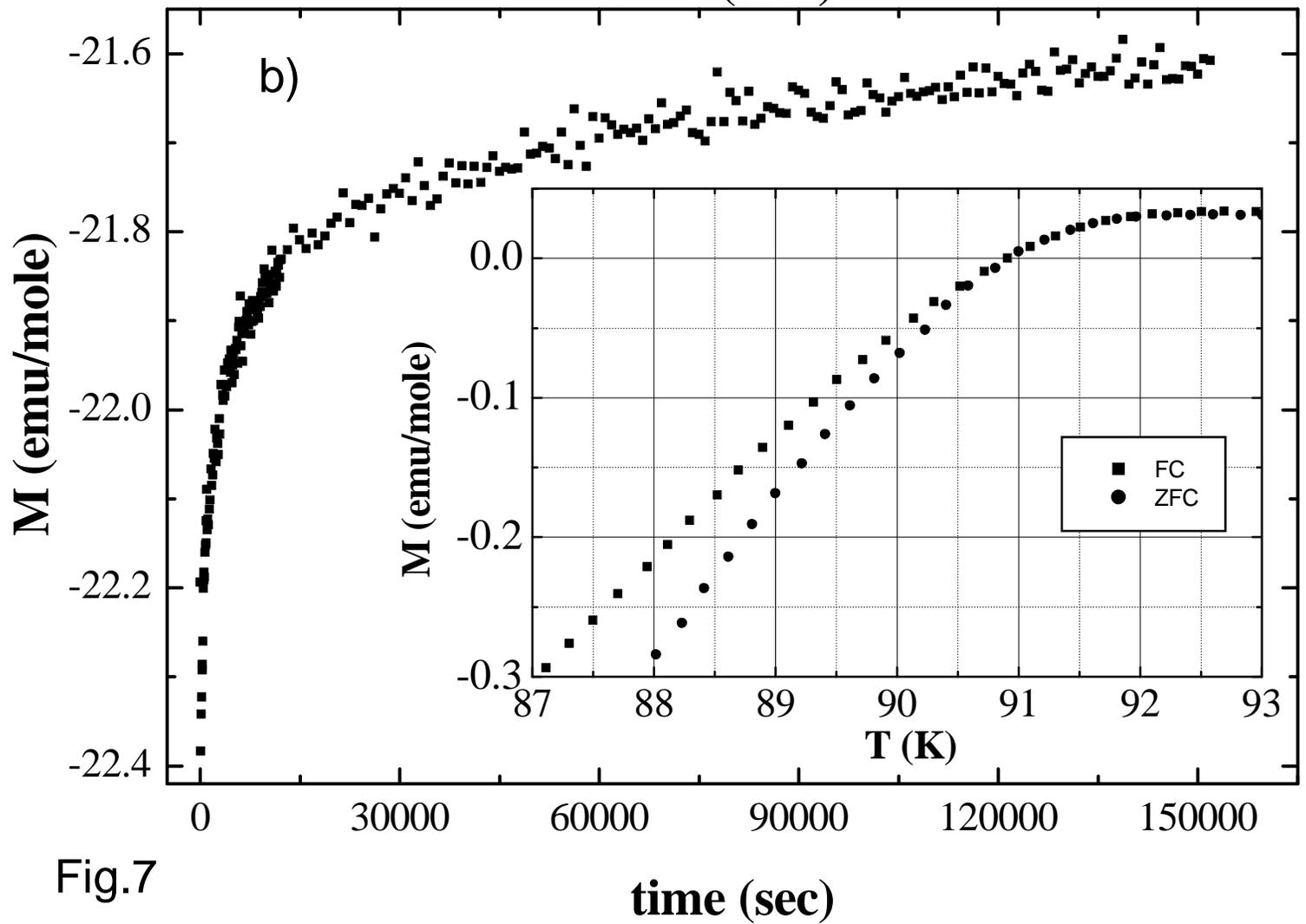
\end{document}